\documentclass[a4paper]{jpconf}
\usepackage{graphicx}
\usepackage{rotating}
\usepackage{epstopdf}
\usepackage[utf8]{inputenc}
\usepackage{amsmath}
\usepackage{amssymb}
\usepackage[T1]{fontenc}
\begin{document}
\title{Rotating and twisting charged black holes with cloud of strings and quintessence}

\author{M F A R Sakti$^1$, H L Prihadi$^1$, A Suroso$^{1,2}$, and F P Zen$^{1,2}$}

\address{$^1$ Theoretical Physics Laboratory, THEPi Division, Institut Teknologi Bandung,
Jl. Ganesha 10 Bandung 40132, Indonesia}
\address{$^2$ Indonesia Center for Theoretical and Mathematical Physics (ICTMP),
Jl. Ganesha 10 Bandung 40132, Indonesia}

\ead{fitrahalfian@gmail.com}

\begin{abstract}
We find a charged spherically symmetric black hole solution with the existence of a cloud of strings and quintessential matter. Then we apply the Demia{\'n}ski-Newman-Janis algorithm to generate the rotating and twisting counterpart. The thermodynamic properties of this black hole solution are further investigated.
\end{abstract}

\section{Introduction}
It is proved that the expansion of our universe is accelerated by a mysterious creature called dark energy. Dark energy may consist of any matter since there is no observational evidence to see its content. Quintessence is one of the matters that can be described as a dark-energy model. It is portrayed as a scalar field that governs the accelerated expansion of the universe where it gives rise to the negative pressure \cite{PerlmutterAstroJ,Copeland2006}. Furthermore, the scalar field is also studied in some gravitational theories on the cosmological scale \cite{Arianto2007,Arianto2008,Zen2009,Arianto2010,Feranie2010,Arianto2011,Suroso2013,Suroso2015,Getbogi2016} and on the locally astrophysical manifestations \cite{SaktiBosonStar,SaktiIntBoson,SaktiQball,SaktiAntoPRD2021}. Interestingly, Kiselev \cite{Kiselev2003} has  studied the scalar field using some arbitrary assumption in a black hole framework. Further studies have been also investigated for some more general cases of black holes in modified gravity \cite{Maharaj2017,Lee2018,Wang2017a, Toshmatov2017,Wang2017b,Ghosh2016,SaktiDarkUni2021}.

One of the theories that tries to merge the gravitational theory and quantum mechanics is the string theory. It tries to explain the whole theories by assuming that the fundamental ingredient of our universe is the one-dimensional strings. Letelier \cite{Letelier1979} has tried to extend the idea of a string to a black hole of which the manifestation of the string's existence might affect its properties. As we know that, rotating black hole is appropriate enough with the observational data that tells that this kind of object exists. In particular, there is no study that investigates the rotating and twisting charged black hole in the presence of a cloud of strings. Although we can find some studies about a cloud of strings within a black hole system in modified gravitational theories, it will be more measurable to consider the cloud of strings for rotating black holes in Einstein gravity. To get more generalized solution, twisting parameter or NUT charge should be considered also.

In this article, we wish to obtain a black hole solution consisting of several matters. To generalize the solution, we consider the existence of quintessence and a cloud of strings as given in \cite{Toledo2018}. In particular, in a special case of the equation of state parameter of the quintessential matter, the contribution of the quintessence is similar with a cloud of strings. The main aim of this article is to investigate the rotating and twisting charged black hole solution with the presence of a cloud of strings and quintessence in general relativity. Even for more general modified gravitational theory, a cloud of string has been studied for example in $ f(R) $ gravity \cite{LeeGhosh2015} and Lovelock theory \cite{Graca2018}.

We organize the remaining parts of the paper as follows. In the section 2, we derive the charged black holes solution with cloud of strings and quintessence. Then the rotating and twisting solution is derived in the section 3 using DJN algorithm. In the section 4, the thermodynamics properties are investigated. Finally, we summarize our article in the last section.

\section{Charged Black Holes with Cloud of Strings and Quintessence}
Before we derive the rotating and twisting charged black hole solution with the presence of a cloud of strings and quintessence. We must derive the spherically symmetric metric solution which will be applied to the Demia{\'n}ski-Janis-Newman algorithm to insert spin and NUT charge. In this section, we wish to derive a charged black holes solution with the cloud of strings and quintessence. However, the more general solution in Lovelock gravity without electromagnetic field has been derived in \cite{Toledo2018}. There are three contributions of matter that we will calculate, i.e. the electromagnetic field,  a cloud of strings, and the quintessence. The elaborated derivation of energy-momentum tensor of a cloud of string can be found in \cite{Letelier1979}. Letelier has generalized the Schwarzschild solution corresponding to a black hole by a spherically symmetric cloud of strings. The Lagrangian density of a cloud of strings is given by
\begin{equation}
\mathcal{L}_s = m\left( -\frac{1}{2}\Sigma^{\mu\nu}\Sigma_{\mu\nu} \right)^{1/2}. \
\end{equation}
$ m $ is just a positive constant related to the string tension. $ \Sigma^{\mu\nu} $ is a bivector defined by
\begin{equation}
 \Sigma^{\mu\nu} = \epsilon^{ab}\frac{\partial x^\mu}{\partial \lambda^a}\frac{\partial x^\nu}{\partial \lambda^b}, \
\end{equation}
where $ \epsilon^{ab} $ is a Levi-Civita tensor, $ \lambda^a (\lambda^a=\lambda^0,\lambda^1) $ is a parameterization of world sheet described by the string with induced metric
\begin{equation}
h_{ab} = g_{\mu\nu}\frac{\partial x^\mu}{\partial \lambda^a}\frac{\partial x^\nu}{\partial \lambda^b}.\
\end{equation}
We need the following identities
\begin{equation}
\Sigma^{\mu[\alpha}\Sigma^{\beta\gamma]}=0,~~~ \nabla_\mu \sigma^{\mu[\alpha}\Sigma^{\beta\gamma]}=0, ~~~\Sigma^{\mu\sigma}\Sigma_{\sigma\tau}\Sigma^{\tau \nu} = \textbf{h}\Sigma^{\nu\mu},\label{eq:strinidentities}
\end{equation}
where $ \textbf{h} $ is the determinant of $ h_{ab} $. Using $ T_{\mu\nu}= 2\partial \mathcal{L}/\partial g_{\mu\nu} $, we can obtain
\begin{equation}
B_{\mu\nu} =\rho_s\frac{\Sigma_{\mu\sigma}\Sigma^\sigma_\nu}{(-\textbf{h})^{1/2}}.
\end{equation}
Moreover, by employing the identities (\ref{eq:strinidentities}), we also find $ \partial_\mu(\sqrt{-g}\rho_s\Sigma^{\mu\nu})=0 $. So, the non-vanishing components of energy-momentum tensor take form
\begin{eqnarray}
B^t_t = B^r_r =- \frac{b}{r^2},~~~
B^\theta_\theta = B^\phi_\phi = 0,\label{eq:stringem}\
\end{eqnarray}
where $ b $ is related to the strength of the cloud. We choose the metric convention ($ -,+,+,+ $). 

Then the appropriate general expression for the energy-momentum tensor of the quintessence is given by \cite{Kiselev2003}
\begin{eqnarray}
&& Q^0_0 =  -\rho_q(r), \nonumber\\
&& Q^i_j =  -3\rho_q(r) \omega_q \left[-(1+3B)\frac{r_i r^j}{r_n r^n} + B \delta^i_j \right].  \
\end{eqnarray}
The spatial part of the energy-momentum tensor is proportional to the time component with the arbitrary parameter $ B $ depending on the internal structure of the quintessence. By taking the isotropic average over the angles, we gain
\begin{eqnarray}
\big \langle Q^i_j \big \rangle = \rho_q(r) \omega_q\delta^i_j = p_q(r) \delta^i_j,
\end{eqnarray}
where we have used $\big \langle r_ir^j \big \rangle = \frac{1}{3}\delta^i_j r_n r^n $. Therefore, we can derive the relation
$p_q = \omega_q \rho_q$. Hence, we find
\begin{equation}
Q^t_t =Q^r_r =- \rho_q(r),~~~
Q^\theta_\theta = Q^\phi_\phi = \frac{1}{2}\rho_q(r)(3\omega_q +1),\label{eq:quinem}
\end{equation}
which implies that $ B = -(3\omega_q +1)/6\omega_q $ and $ \omega_q $ is the equation of state parameter. This parameter plays a role to determine the perfect-fluid matter domination as in the scenario of accelerated expansion of the universe. We may impose $ -1 < \omega_q < -1/3 $ to obtain quintessential dark energy domination. However, $ \omega_q $ may vary for any value since there is no constraint in the black hole solution.

The last matter component is the electromagnetic field. It is a very common component in the black hole solution. We will not consider any special Langrangian density for electromagnetic part. So, the general energy-momentum form of this matter tensor are
\begin{equation}
E_{\mu \nu} = 2\left(F^\alpha_\mu F_{\nu \alpha} -\frac{1}{4}g_{\mu \nu}F^{\alpha \beta}F_{\alpha \beta} \right).
\end{equation}
Using Bianchi identity $ F_{[\mu \nu ;\gamma]} $ and Maxwell equation $ \nabla_\mu F^{\mu \nu}=0 $ then assuming non-zero components $ F^{01}, F^{23} $, we can find
\begin{equation}
E^t_t = E^r_r =- \frac{e^2+g^2}{r^4},~~~
E^\theta_\theta = E^\phi_\phi = \frac{e^2+g^2}{r^4},\label{eq:emem}
\end{equation}
where $ e $ and $ g $ are the electric and magnetic charges, respectively. The electromagnetic potential of this matter tensor is
\begin{eqnarray}
A_\mu dx^\mu = -\frac{e}{r}dt+g\cos\theta d\phi .\
\end{eqnarray}
Its derivation can be found in \cite{SaktiCQG}.

We will use this metric ansatz
\begin{equation}
ds^2 = -f(r)dt^2 +f(r)^{-1}dr^2 +r^2 d\Omega^2,\label{eq:spersym}
\end{equation}
where $f (r)$ is a function depending on radial coordinate only. We also have two-dimensional sphere as $d \Omega^2 = d\theta^2 + \sin^2 \theta d\phi^2 $. By employing Einstein field equation $ G_{\mu\nu} = T_{\mu\nu} $ and all of the matters (\ref{eq:stringem}), (\ref{eq:quinem}) and (\ref{eq:emem}) are considered, we find the following set of equations
\begin{equation}
\frac{1}{r^2}(f'r-1+f)=-\rho_q-\frac{e^2}{r^4}-\frac{g^2}{r^4}-\frac{b}{r^2}, \
\end{equation}
\begin{equation}
\frac{1}{r^2}(f'r +\frac{1}{2}f''r^2)=\frac{1}{2}\rho_q(3\omega_q +1)+\frac{e^2}{r^4}+\frac{g^2}{r^4}. \
\end{equation}
Solving the above equations, the function $f (r)$ reads as
\begin{equation}
f(r) = 1- \frac{2M}{r}+\frac{q^2}{r^2}- \alpha r^{-1-3\omega_q}-b .
\end{equation}
We have written $e^2 +g^2 =q^2$. $M$ and $ \alpha$ are defined as black hole's mass and quintessential intensity, respectively.
For vanishing charges, it has been derived in \cite{Toledo2018} by turning off the parameters related to non-linear Ricci scalar term. We also have
\begin{equation}
\rho_q = \frac{3\alpha \omega_q}{2r^{1+\omega_q}}. \
\end{equation}
We already find the charged spherically symmetric black hole solution with the presence of a cloud of strings and quintessence. All properties of this solution are determined by $ M, e, g, \alpha, \omega_q $ and $  b $. In the next section, we will derive the rotating and twisting counterpart of this black hole solution. The Demia{\'n}ski-Janis-Newman (DJN) algorithm will be implemented to derive the solution.

\section{Rotating and Twisting Black Hole Solution}

The rotating and twisting solution of a charged spherically symmetric black hole with the presence of a cloud of strings and quintessence has not been derived. We profit to use the DJN algorithm to obtain that solution. Earlier method of this algorithm is shown in \cite{Demianski1972} using tetrad formalism. However, Erbin simplifies this by just using some simple complex coordinate transformations \cite{Erbin2015,Erbin2016} which we will employ within this section.  We also study the NUT black hole in several different cases \cite{SaktiIJMPD2018,SaktiConfMicEntro,SaktiEPJP,Saktideformed2019,SaktiCQG,SaktiNucPhysB2020,HadyanIJMPD}. For some physical interpretation of the NUT charge, we can find the explanation in \cite{Badawi2006,Nouri1997}.

Starting the algorithm, the spherically symmetric metric (\ref{eq:spersym}) can be written in the form
\begin{eqnarray}
ds^2=-f_t(r)dt^2+f_r(r)dr^2+ f_s(r) \left[d\theta^2 +H^2(\theta)d\phi^2 \right]. \label{eq:generalsphericallysymmetric}\
\end{eqnarray}
The function $ H(\theta) $ is defined as
\begin{equation}
    H(\theta) = \left\{\begin{array}{lr}
        \sin\theta, ~~ k=1, \\
        1, ~~~~~~ k=0,\\
        \sinh\theta, k=-1,
        \end{array}\right.
\end{equation}
where $ k $ is the sign of the surface curvature. Using a null coordinate transformation $ dt = du + \sqrt{\frac{f_r}{f_t}}dr $, it makes the metric (\ref{eq:generalsphericallysymmetric}) transforms as
\begin{eqnarray}
ds^2 = -f_t du^2 -2\sqrt{f_t f_r} du dr + f_s \left(d\theta^2 +H^2 d\phi^2 \right) . 
\end{eqnarray}
After that, we need to use the complex coordinate transformation defined as
\begin{equation}
r = \hat{r} + i F(\theta), ~~ u = \hat{u} + i G(\theta), \label{eq:r,u}
\end{equation}
where $ \hat{u},\hat{r} $ are real while $ F(\theta), G(\theta) $ are two arbitrary functions depending on $ \theta $. To find the twisting and rotating solution, $ F(\theta)$ and $ G(\theta) $ are given by
\begin{eqnarray}
 F(\theta) = -n - a \cos\theta, ~~~~~
 G(\theta) = a \cos\theta + 2n \textrm{ln}(\sin\theta) -2n \text{ln}\left(\tan \frac{\theta}{2} \right), \label{eq:FGfunction}
\end{eqnarray}
where $ a $ and $ n $ are the spin and NUT charge, respectively. The complexification needs to do also to the coordinate $ r $ and mass $ M $ that read as 
\begin{equation}
\hat{r} \rightarrow  \frac{1}{2}(\hat{r}+\bar{r}) = \textrm{Re}(\hat{r}), ~~~~
\hat{r}^2 \rightarrow  |\hat{r}|^2, \nonumber\
\end{equation}
\begin{equation}
\frac{M}{\hat{r}} \rightarrow \frac{1}{2}\left( \frac{M}{\hat{r}}+\frac{\bar{M}}{\bar{r}} \right) = \frac{M\bar{r}+\bar{M}\hat{r}}{|\hat{r}|^2}, ~~~M \rightarrow M+in, \
\end{equation}
Note that $ \bar{r} $ is the conjugate of $ \hat{r} $ and  $ \bar{M} $ is the conjugate of $ M $.  Furthermore, we need to use Giampieri's ansatz \cite{Giampieri1990} on the angular coordinate, i.e. $ id\theta = H(\theta) d\phi $. Hence, Eq. (\ref{eq:r,u}) is transformed as
\begin{eqnarray}
dr = d\hat{r} + F'(\theta)H(\theta) d\phi, ~~~~~
du = d\hat{u} +  G'(\theta)H(\theta) d\phi . \label{eq:dr1,du1}\
\end{eqnarray}

Finally, we can replace $ f_t, f_r, f_s $ by $ \bar{f}_t, \bar{f}_r, \bar{f}_s $ because we already use the
new complex coordinates. Then the twisting and rotating solution in Eddington-Finkelstein
coordinates is given as follows
\begin{eqnarray}
ds^2 &=& - 2\sqrt{\bar{f}_t \bar{f}_r}(d\hat{u} d\hat{r}+G'H d\hat{r}d\phi) -2(\bar{f}_t G'H  +\sqrt{\bar{f}_t \bar{f}_r}F'H)d\hat{u}d\phi \nonumber\\
&-&\bar{f}_t d\hat{u}^2+\bar{f}_s d\theta^2 -(\bar{f}_tG'^2 H^2 + 2\sqrt{\bar{f}_t \bar{f}_r}F'G'H^2 -\bar{f}_s H^2)d\phi^2 .\label{eq:nutspinkriskalgen}
\end{eqnarray}
In order to find the solution in Boyer-Linquist coordinates, we need to transform metric (\ref{eq:nutspinkriskalgen}) with the following coordinates transformation defined as
\begin{eqnarray}
d\hat{u} = d\hat{t} - g(\hat{r}) d\hat{r}, ~~~~~
 d\phi = d\varphi - h(\hat{r}) d\hat{r}, \label{eq:integrabletheta}
\end{eqnarray} 
where
\begin{eqnarray}
g(\hat{r}) = \frac{(\bar{f}_t \bar{f}_r)^{-\frac{1}{2}}\bar{f}_s - F'G'}{\frac{\bar{f}_s}{\bar{f}_r}+F'^2} =\frac{r^2 +(n+a)^2}{\Delta}, ~~~
h(\hat{r}) = \frac{F'}{H\left(\frac{\bar{f}_s}{\bar{f}_r}+F'^2\right)} =\frac{a}{\Delta}.\label{eq:integrable}
\end{eqnarray}
Note that the functions $ g(\hat{r}),h(\hat{r}) $ cannot depend on coordinate $ \theta $ because it can make the
transformations (\ref{eq:integrabletheta}) unintegrable \cite{Erbin2015,Erbin2016}. Moreover, using (\ref{eq:integrabletheta}) will result in the general twisted
rotating metric in Boyer-Linquist coordinates reads as
\begin{eqnarray}
ds^2 &=& -\frac{\Delta}{\rho^2}\left[dt - \bigg\{ a\sin^2\theta +2n(1-\cos\theta) \bigg\} d\varphi \right]^2+ \frac{\rho^2}{\Delta}dr^2 \nonumber\\
& &+\rho ^2 d\theta ^2 +\frac{\sin^2\theta}{\rho ^2}\left[adt- \bigg\{r^2+(a+n)^2 \bigg\}d\varphi \right]^2 , \label{eq:metricresult1}
\end{eqnarray}
where $\rho^2 = r^2 + (n+a\cos\theta)^2 $, $ \Delta = r^2 -2Mr +a^2+q^2-n^2-\alpha r^{\upsilon}-br^2 $ and $ \upsilon=1-3\omega_q $. We already omit the hat $ (\hat{~}) $ on $ t,r $. For vanishing $ b $, our solution is similar as the solution in \cite{SaktiCQG} when $ \kappa\lambda =0 $. Our solution contains a cloud of strings and quintessence that has a similar contribution in the metric when $ \omega_q = -1/3 $ for the perfect-fluid dark matter domination.

\section{Thermodynamic Properties}

\begin{figure}
\resizebox{1.0\textwidth}{!}{
\includegraphics[scale=1]{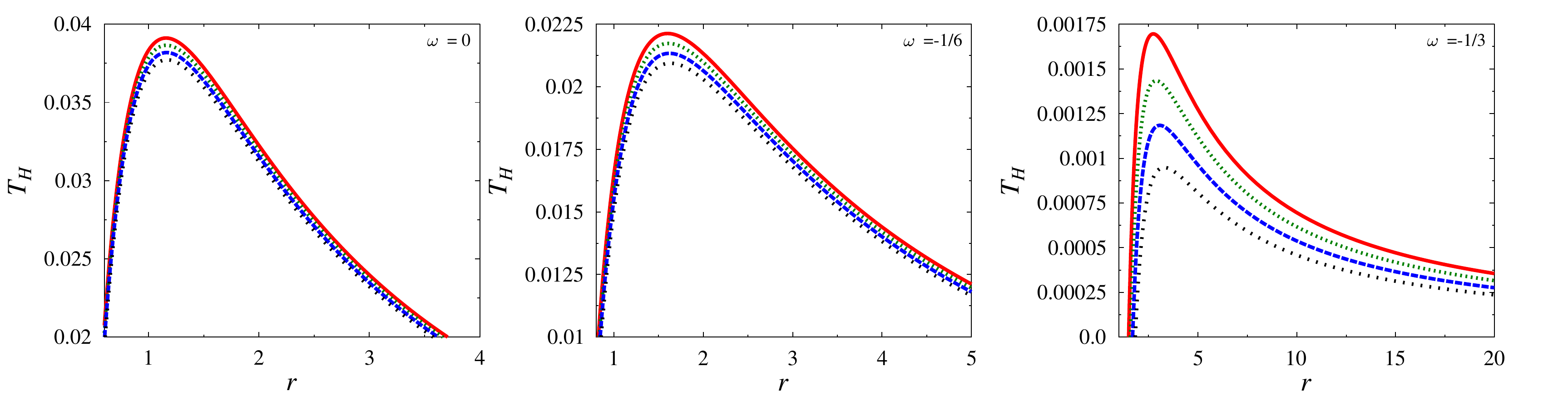}
}
\caption{Plot of Hawking temperature. We take $a = 0.5$, $n = 0.3$, $ \alpha =0.9 $, $q = 0.2$, $M = 1$ and $b = 0.01, 0.02, 0.03, 0.04$ for red, green, blue, purple lines, respectively.}
\label{fig:temperature}
\end{figure}

In this section, we investigate some important thermodynamic quantities. For the black hole
solution (\ref{eq:metricresult1}), the number of horizon is dependent of $ \omega_q $. To study the thermodynamics of this
solution, we need to know the location of the event horizon. Hence, here we will define the
thermodynamics in term of event horizon $ r_+ $. To calculate the Hawking temperature, we may
govern the tunneling method \cite{Parikh2000,Banerjee2008,Ma2008}. By governing this method, the metric is allowed to be
diagonal. Then $ d\theta=d\phi=0 $ is taken and the angular coordinate is chosen to be $ \theta=0 $. Hence,
the metric and Hawking temperature can be expressed respectively as
\begin{equation}
ds^2 =-f(r)dt^2 +f(r)^{-1}dr^2, \
\end{equation}
\begin{equation}
T_H = \frac{\partial_r f(r)}{4\pi}\bigg|_{r=r_+}.
\end{equation}
After the tunneling method is applied, the metric now can be read as
\begin{equation}
ds^2 = -\frac{\Delta}{r^2 +(n+a)^2}dt^2 + \frac{r^2 +(n+a)^2}{\Delta}dr^2 .\ 
\end{equation}
Hence the Hawking temperature can be computed and straightforwardly produces
\begin{equation}
T_H = \frac{2(r_+ - M)-br_+ -\alpha (1-3\omega_q)r_+^{-3\omega_q}}{4\pi [r_+^2 + (n+a)^2]} . \label{eq:Hawkingstring}
\end{equation}
We show the plot of the Hawking temperature for some specific value of parameters as shown in figure \ref{fig:temperature}. As we can see from figure \ref{fig:temperature}, the maximum Hawking temperature is smaller for the smaller value of $ \omega_q $. When the parameter $ b $ is varied, it gives higher Hawking temperature for smaller $ b $. So the presence of a cloud of string will reduce the Hawking temperature.

The angular momentum of solution (\ref{eq:metricresult1}) is then
\begin{equation}
\Omega_H = - \frac{g_{03}}{g_{33}}\bigg|_{r=r_+} = \frac{a}{r^2_+ + (n+a)^2}. \
\end{equation}
Note that the quintessence causes another thermodynamic variable to appear \cite{Chen2008,Sekiwa2006}
\begin{equation}
\Theta_H =\frac{\partial M}{\partial \alpha}\bigg|_{r=r_+}=-\frac{1}{2}r_+^{\upsilon -1},
\end{equation}
Besides, the cloud of strings may also give a contribution to the thermodynamics.  Hence, the new thermodynamic variable is given by
\begin{equation}
B_H =\frac{\partial M}{\partial b}\bigg|_{r=r_+}=-\frac{1}{2}r_+ . \
\end{equation}
$ B_H $ is computed when $ S,J,Q_e,Q_g, \alpha $ are
held constant. These generalized forces are needed to satisfy the first law of black hole's thermodynamics \cite{Chen2008,Sekiwa2006}. As it can be seen that $ \omega_q = -1/3 $ will produce $ \Theta_H = B_H $. Note that the quintessence and a cloud of strings effect is implicitly contained in $ r_+ $, even we can see obviously the effect of it in Hawking temperature. It will be fascinating for further study to investigate this condition in microscopical scale.

The area of the black hole can be calculated as
\begin{eqnarray}
A_{BH} &=& \int \sqrt{g_{22} g_{33}} d\theta d\varphi = 4\pi [r_+^2 + (a+n)^2]. \label{eq:areabh} \
\end{eqnarray}
By employing the area law of the black hole, the Bekenstein-Hawking entropy can be computed as
\begin{eqnarray}
S_{BH} = \frac{A_{BH}}{4} = \pi [r_+^2 + (a+n)^2]. \label{eq:bhentropy}
\end{eqnarray}
When $ \omega_q = -1/3 $, $ r_+ $ is proportional to $ 1/(1-b -\alpha)^2 $ that we can calculate from $ \Delta=0 $. Since $ r_+ $ is proportional to $ 1/(1- b - \alpha)^2 $, the value of $ b + \alpha $ is restricted as $(b+\alpha) \neq 1 $ in order to generate non-singular entropy. For vasnishing a cloud of strings, it has been found in \cite{SaktiCQG}. For constant value of $ J,Q_e,Q_g,\alpha, b $, we can also obtain the heat capacity using the relation which is given by \cite{Maharaj2017,Lee2018,Cai2004}
\begin{eqnarray}
C = \frac{dM}{dT_H}\bigg|_{r=r_+} = \frac{dM/dr_+}{dT_H/dr_+}\bigg|_{r=r_+}, \label{eq:heatcap}\
\end{eqnarray}
where the mass can be defined from $ \Delta=0 $ as
\begin{equation}
M = \frac{1}{2r_+}\left(r_+^2 -\alpha r_+^{1-3\omega_q} + n^2 +a^2 +q^2 -br_+^2 \right) .\
\end{equation}
Hence from Eq. (\ref{eq:heatcap}), we find that
\begin{equation}
C = \frac{2 \pi [r_+^2 + (a + n)^2]^2 (-a^2 + n^2 - q^2 + r_+^2 - b r_+^2 + 3 \alpha\omega_q r_+^{1 - 3 \omega_q} )}{r^2 (r_a + r_b )} , \label{eq:heatcap1}\
\end{equation}
where
\begin{eqnarray}
r_a &=& 2 (a^2 - n^2 + q^2 - r_+^2 + b r_+^2 - 3 r_+^{1 - 3 \omega_q} \omega_q \alpha), \nonumber\\
r_b &=& \{r_+^2 + (a + n)^2\} \{2 - 2 b + 3 \alpha\omega_q(1 - 3 \omega_q) r_+^{1 - 3 \omega_q}\} \nonumber.
\end{eqnarray}

\begin{figure}
\resizebox{1.0\textwidth}{!}{
\includegraphics[scale=1]{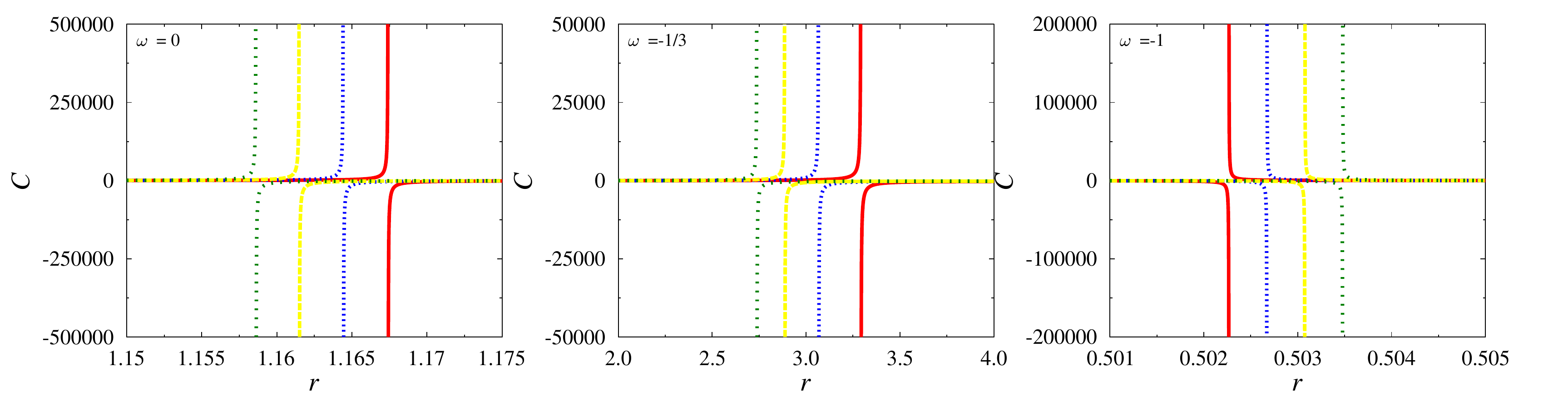}
}
\caption{Plot of heat capacity. We take $a = 0.5$, $n = 0.3$, $ \alpha =0.9 $, $q = 0.2$, and $b = 0.01, 0.02, 0.03, 0.04$ for red, blue, yellow, green lines, respectively.}
\label{fig:heatcapacity}
\end{figure}
The dependence of $ \alpha, \omega,b $ on area and entropy is implicitly contained on $ r_+ $ while for the heat capacity, it is obviously seen in Eq. (\ref{eq:heatcap1}). Note that the thermodynamic stability of any system is related to the sign of the heat capacity. As we are talking about a black hole system, when the heat capacity is positive $ (C>0) $, the black hole will be stable thermodynamically. Whereas when the heat capacity is negative $ (C<0) $, the black holes are said to be unstable. Spin and NUT charge also play an important role on those thermodynamic variables. In the entropy, the square of spin and NUT charge are proportional to the entropy. Besides, in the heat capacity, it is not really obvious but those parameters still possess a similar contribution as given in figure \ref{fig:heatcapacity}.

\section{Summary}

We have obtained the rotating and twisting charged black hole solution with the presence of a cloud of strings and quintessence using DJN algorithm. The thermodynamics has also been also studied. The Hawking temperature reduces due to the presence of the quintessence and a cloud of strings. The similar effect also occurs on the heat capacity. The perfect-fluid domination $ \omega_q = -1/3 $ leads us the special condition which is similar with the cloud of strings. It also produces that $ \Theta_H = B_H $ where those two variables are related to each matter besides the electromagnetic field. One more interesting event occurs also for the entropy because there is a restriction of the value of $ \alpha $ and $ b $ in order to describe the non-singular entropy. Hence, our solution can provide a correspondence between the string theory and the gravitational theory. Further study about the microscopic description of this entropy should be investigated.


\section*{Acknowledgments}
We gratefully acknowledge support by Riset PMDSU 2018 from Ministry of Research, Technology, and Higher Education of the Republic of Indonesia. M F A R S also thanks all members of Theoretical Physics Laboratory, Institut Teknologi Bandung for the valuable support.

\end{document}